# Giant modulation of the electronic band gap of carbon nanotubes by dielectric screening

*Lee Aspitarte[1], Daniel R. McCulley[1], Andrea Bertoni[2], Joshua O. Island[3], Marvin Ostermann[3], Massimo Rontani[2], Gary A. Steele[3] and Ethan D. Minot[1*]*

[1] Department of Physics, Oregon State University, Corvallis, OR, 97331, USA

[2] Istituto Nanoscienze-CNR, Via Campi 213a, I-41125 Modena, Italy

[3] Kavli Institute of Nanoscience, Delft University of Technology, Lorentzweg 1, Delft 2628 CJ, The Netherlands

[*]Corresponding author email: ethan.minot@oregonstate.edu

Carbon nanotubes (CNTs) are a promising material for high-performance electronics beyond silicon. But unlike silicon, the nature of the transport band gap in CNTs is not fully understood. The transport gap in CNTs is predicted to be strongly driven by electron-electron (e-e) interactions and correlations, even at room temperature. Here, we use dielectric liquids to screen e-e interactions in individual suspended ultra-clean CNTs. Using multiple techniques, the transport gap is measured as dielectric screening is increased. Changing the dielectric environment from air to isopropanol, we observe a 25% reduction in the transport gap of semiconducting CNTs, and a 32% reduction in the band gap of narrow-gap CNTs. Additional measurements are reported in dielectric oils. Our results elucidate the nature of the transport gap in CNTs, and show that dielectric environment offers a mechanism for significant control over the transport band gap.

Carbon nanotubes (CNTs) are a promising platform to move integrated-circuit technology beyond the current limits of silicon.[1–3] However, there are critical open questions regarding the nature of the transport band gap in CNTs, and in particular the role that electron interactions may play in determining this band gap. In quantum transport experiments,[4] electron-electron (e-e) interactions lead to phenomenon such as Luttinger liquid physics[5,6] and Wigner crystal formation in CNTs,[7,8] and may also explain anomalous spin-orbit coupling in CNTs,[9] and anomalous band gaps in nominally-metallic CNTs.[10] Beyond this low-energy physics typically observed at cryogenic temperatures, theoretical models suggest that e-e interactions play a significant role in nanotube electronic properties even at room temperature. A giant renormalization of the single-particle gap (the transport gap) is predicted. The key experimental signature of this effect is the relationship between the transport gap and the dielectric environment (Fig. 1a).[11]

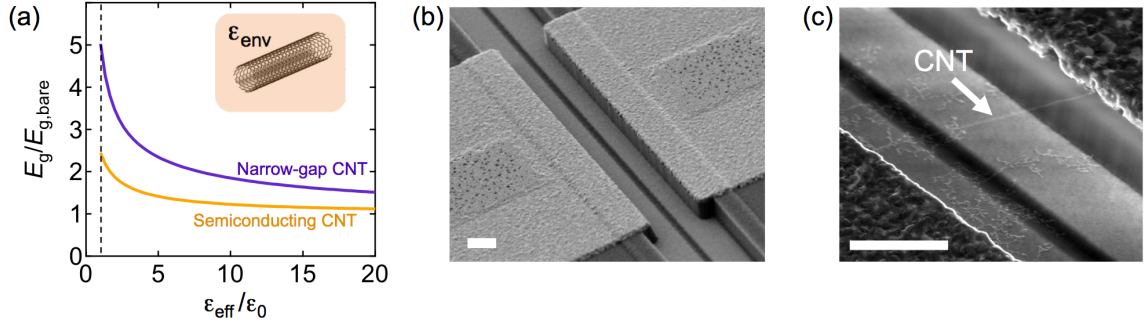

**Figure 1.** (a) Theoretical calculation of the enhancement of the transport gap, $E_g/E_{g,bare}$, as a function of the effective dielectric constant, $\epsilon_{eff}$, for a narrow gap and semiconducting CNT, both with diameter $D = 2$ nm. In experiment, $\epsilon_{eff}$ is controlled by the dielectric environment. The transport gap calculation is based on the random-phase approximation method,[11] and includes a contribution due to curvature (see Eqn. 2). Similar enhancement factors are found when $D = 1$ nm (see Supporting Information). (b) Electron microscopy image of the device structure. The source and drain electrodes are separated by 2 μm. The scale bar is 1 μm. Catalyst is deposited on top of the electrodes (4-μm square areas). The two gate electrodes at the bottom of the trench are separated by a 250 nm gap. The gate electrodes are electrically connected (off chip) and operated as a single gate. (c) A suspended CNT bridging the gap between source and drain electrodes. The scale bar is 1 μm.

A screening-dependent transport gap is particularly significant because a mutable gap would dramatically affect the design of nanoelectronic devices. For CNTs, there is preliminary experimental evidence that such an effect exists. A scanning tunneling microscopy study showed that a semiconducting CNT in direct contact with the metal substrate had a 25% smaller transport gap than a similar CNT that, by chance, was positioned a few angstroms above the metal substrate.[12] In other experiments, *I-V* curves of pn junctions made from semiconducting CNTs have been interpreted in a framework of band gap renormalization.[13,14] Screening-induced changes in transport gap have been confirmed in other low-dimensional nanomaterials. Recent experiments showed that single-layer MoSe$_2$ samples prepared on graphite had 11% smaller single-particle gap than single-layer MoSe$_2$ prepared on bilayer graphene.[15] In our current work, we demonstrate changes in the CNT transport gap greater than 30%, an extraordinarily large effect.

A theoretical framework for understanding the influence of dielectric environment on the transport gap of semiconducting CNTs was first developed by Ando.[11] This framework begins with the non-interacting model for CNT band structure,[4] in which the "bare" transport gap of semiconducting CNTs is given by

$$E_{g,bare} \approx 0.7 \text{ eVnm}/D, \qquad (1)$$

where *D* is CNT diameter. Ando predicted that this bare transport gap is enhanced by a factor of ~ 2 when e-e interactions are considered. The enhancement factor depends on the dielectric constant of environment surrounding the CNT, as illustrated in Fig. 1a. Ando predicted that the screening dependence of the renormalized gap, $E_g$, would not be revealed by single-photon optical absorption/emission resonances because exciton binding energy also changes with dielectric screening. While Ando's theory describing $E_g$ in semiconducting CNTs has been corroborated by additional theoretical work,[16–21] experimental verification of the relationship between $E_g$ and dielectric environment has remained lacking.

In this work we also investigate the transport gap in narrow-gap CNTs (often called "metallic" CNTs). Carbon nanotubes that are considered metallic based on a simple zone-folding approximation, have a small curvature-induced gap. Without considering e-e interactions, the bare curvature-induced transport gap is calculated to be,[4]

$$E_{g,bare} \approx (50 \text{ meV} \cdot \text{nm}^2/D^2)\cos 3\theta, \quad (2)$$

where $\theta$ is chiral angle. We have extended Ando's theory to predict the enhancement of this curvature-induced transport gap in narrow-gap CNTs (Fig. 1a).

To test the predicted relationship between $E_g$ and dielectric environment we developed in-situ methods to measure individual CNTs in various dielectric environments. Using individual ultra-clean suspended CNTs (Fig. 1b and c), we perform measurements before and after submersion in dielectric liquids. Multiple techniques are used to determine the modulation of the band gap and demonstrate the extraordinary sensitivity of CNTs to dielectric screening.

## Results

Ultra-clean suspended CNTs were fabricated by growing CNTs over pre-made electrode structures.[22] Details of the device fabrication are given in the methods section. Because CNT growth is the final fabrication step, the CNTs are never exposed to fabrication chemicals, thereby retaining pristine material qualities. After post-selection for CNTs with good electrical conductivity, we have counted 110 narrow-gap CNTs and 187 semiconducting (wide-gap) CNTs over the last 3 years of experiments. Figure 1c shows a completed device.

Figure 2a shows the measurement circuit that was used to characterize transport properties in different dielectric environments. The gate electrodes are held at potential, $V_g$, relative to the drain electrode. The CNT is p-doped when $V_g < 0$ and n-doped when $V_g > 0$. Figure 2b shows an electrostatics simulation, and self-consistent band-bending diagram for a semiconducting CNT gated with $V_g = 3$ V. The source-drain bias is held constant, $V_{sd} = 25$ mV, and the current $I$ is recorded with a low noise current preamplifier (SRS 570). The conductance of the CNT is $G = I/V_{sd}$.

Figures 2c and 2d show measurements of $G$ at room temperature as a function of $V_g$, for a narrow-gap and a semiconducting CNT respectively. The $G(V_g)$ curves were first measured in air. The devices were then submerged in refractive index oil ($n = 1.46$, Cargille) and measured again. This measurement has been repeated on 4 additional narrow-gap CNTs, and 6 additional wide-gap CNTs. In the narrow-gap CNTs, the oil environment reproducibly increases $G$ and the maximum value of the normalized transconductance, where normalized transconductance is defined as $dG/dV_g$. In the semiconducting CNTs, there is no conductance at $V_g > 0$, due to the Schottky barriers for n-type transport (see the band bending diagram in Fig. 2b, and device simulations in the Supporting Information). When $V_g < 0$ the conductance turns on very abruptly, both in air and oil environment.

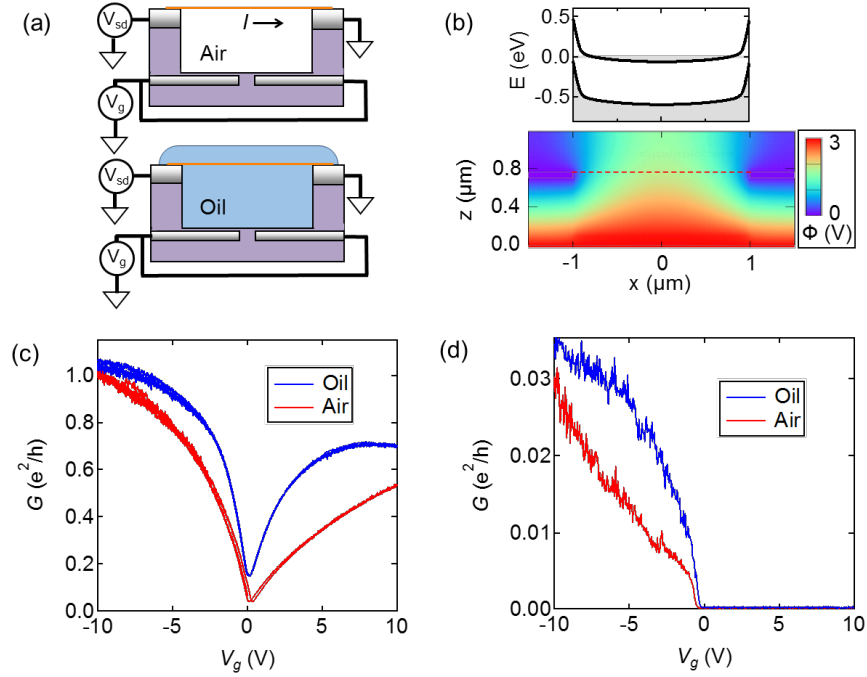

**Figure 2.** a) Schematic of a CNT field-effect transistor in air and oil environments. b) Self-consistent electrostatics calculation of a wide-gap CNT device with $E_g$ = 0.6 eV, $V_g$ = 3 V and an air environment. Top: energy band profile along the CNT. Bottom: spatial map of the electrostatic potential $\Phi(x, z)$. c) Conductance of a narrow-gap CNT in air and oil environments. d) Conductance of a wide-bandgap CNT in air and oil environments.

We first comment on the increase in peak transconductance in narrow-gap CNTs. When the device is submerged in oil, the oil increases the capacitive coupling between the back gate and the CNT. This increased gate capacitance means that more charge carriers are pulled into the CNT for a given $V_g$, consistent with our observations.

The increase in transconductance is one of several observations showing that the dielectric liquid does not introduce electrostatic disorder along the length of the CNT. Electrostatic disorder was a confounding factor in previous attempts to manipulate the dielectric environment of electrically-contacted CNTs. For example, Amer et al. compared pairs of CNT devices made from a single narrow-gap CNT, with the CNT either suspended or touching a SiO$_2$ substrate.[23] In these experiments with a solid dielectric material, the dielectric caused a reduction in transconductance; the high conductance wings in the $G(V_g)$ curve were suppressed, and the low-conductance dip was raised. This flattening of the $G(V_g)$ curve was a clear signature of electrostatic disorder. In contrast, when liquid is added to narrow-gap devices, transconductance is increased (Fig. 2c). We also note the sharp turn-on observed in semiconducting CNT devices both before and after submersion in liquid (Fig. 2d). There is no evidence that the liquid causes a disorder tail in the transistor curve. This is a key insight. Dielectric liquids unlock the possibility of exploring the effect of dielectric screening without the complicating factor of electrostatic disorder.

The remainder of this paper focuses on quantifying the environmentally-induced change in $E_g$ in both narrow-gap and semiconducting CNTs.

**Narrow-gap CNTs**

The overall change in $G$ (Fig. 2c) suggests either a change in $E_g$, or a change in the work function of the metal electrode. We first discuss the role of the metal work function and eliminate the possibility that a change in work function can explain Fig. 2c.

We investigated the influence of the metal work function on device characteristics by modifying the surface adsorbates on the metal electrodes. Derycke et al. showed that a vacuum environment (clean metal surfaces) favors alignment between the metal's Fermi level and the conduction band of the CNT. In contrast, an ambient environment (water and oxygen adsorbates) favors alignment between the metal's Fermi level and the valence band of the CNT.[24] Figure 3 shows $R(V_g)$ curves that were obtained as a narrow gap CNT is brought from a vacuum environment into an ambient environment. In vacuum, the n-type resistance is lower than the p-type resistance, consistent with Fermi level alignment with the conduction band. After exposure to ambient environment, the p-type resistance is lower than n-type resistance. The biggest resistance peak is obtained while the device is transitioning from the vacuum to the ambient environment and $R(V_g)$ is perfectly symmetric (purple curve). While the $R(V_g)$ curve is clearly affected by metal work function (red, purple and black curves), submerging the device in oil (gray curve) has a much larger effect than can be explained by changes in the metal work function. We conclude that changes in metal work function can vary the resistance peak by ±10%, but cannot explain the dramatic reduction in resistance caused by the oil environment.

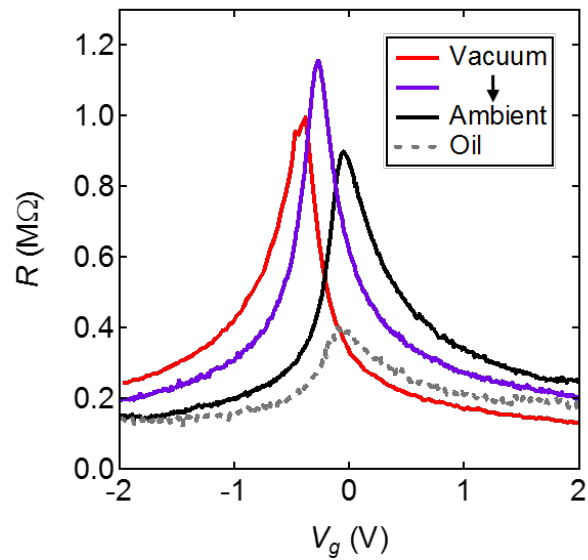

**Figure 3.** The resistance of a narrow-gap CNT measured with different band alignment conditions (red, purple and black curves) and the same CNT measured in refractive index oil (dashed gray curve).

To quantify $E_g$ before and after submersion in oil, we require a transport model that relates the $R(V_g)$ curve to $E_g$. We have taken an empirical approach to establishing such a relationship. We performed Coulomb blockade spectroscopy measurements to determine $E_g$ for a set of 10 narrow-gap CNT devices. For each device we also measured $R(V_g)$ at room temperature. From this data set we find a strong correlation between $E_g$ and the room temperature $R(V_g)$, as described below.

Coulomb blockade spectroscopy was performed in a vacuum environment at $T = 2$ K. The transport gap is resolved by measuring $I$ while varying $V_{sd}$ and $V_g$, as shown in Fig. 4a and c. Figure 4b and d show the corresponding room-temperature measurements of $R(V_g)$ (air environment). The resistance peak is quantified using the parameter $\Delta R = R_{peak} - R_c$, where $R_c$ is the contact resistance of the device. The contact resistance is determined by extrapolating the wings of the $R(V_g)$ curve, which asymptotically approach $R_c$ in an exponential fashion (see SI).

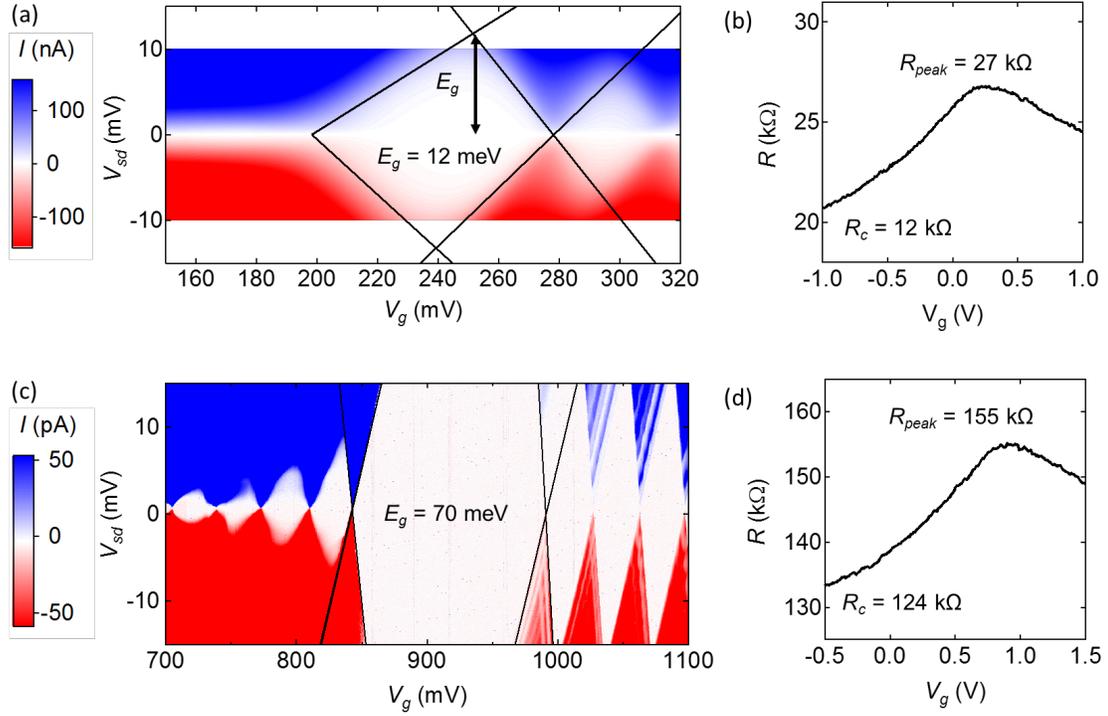

**Figure 4.** Low-temperature and room-temperature transport for two narrow-gap devices. (a) and (c) show CB spectroscopy taken at $T = 2$ K. The lines follow the edges of the $0^{th}$ diamond and are extrapolated to determine the bandgap. (b) and (d) show room temperature resistance as a function of gate voltage.

Measurements of $E_g$ and $\Delta R$ from 10 different narrow-gap CNTs are summarized in Figure 5a. A linear regression of $\ln(\Delta R)$ vs. $E_g$ yields

$$E_g = [58.9 \text{ meV}] \cdot \ln\left(\frac{\Delta R}{[13.3 \text{ k}\Omega]}\right). \qquad (3)$$

If $\Delta R$ measurements are used to estimate $E_g$ (using equation 3), the average residual error is 13 meV.

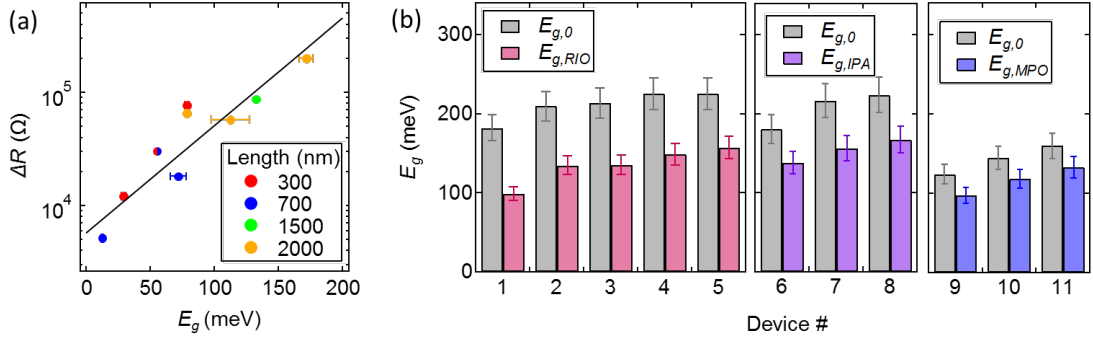

**Figure 5.** (a) Summary of Δ$R$ and $E_g$ values for 10 different narrow-gap CNT devices. The source-drain separation distance, $L$, varies from 300 nm to 2000 nm. (b) The transport gap, $E_g$, measured before and after submersion in refractive index oil (RIO), machine pump oil (MPO), and pure isopropanol (IPA). $E_g$ is calculated from Eqn. 3 using measurements of Δ$R$. The devices are different from those presented in Figure 5a, and all have $L = 2$ μm, except for device 9 which has $L = 1.5$ μm. The estimated experimental error in determination of $E_g$ is ±5% as determined from the measured variation in $E_g$ with band pinning in Figure 3.

Surprisingly, our empirical fit is not improved by accounting for the length of the CNT channel. Insensitivity to channel length may be explained by considering band bending. Only a short segment of the CNT is maximally depleted of charge carriers. The high-resistance segment can be treated as a ballistic channel if the segment length is a few hundred nanometers in length (i.e. less than the scattering length in the narrow-gap CNT).

Using Eqn. 3 we quantify the reduction in $E_g$ when narrow-gap CNTs are submerged in refractive index oil, isopropanol, and machine pump oil (Hyvac, P8900-1) (Figure 5b).

The average reduction in $E_g$ is 39% for refractive index oil, 32% for isopropanol and 28% for machine pump oil.

**Semiconducting CNTs**

We now turn to semiconducting CNTs for which $E_g \gg k_B T$. In these devices the peak resistance in the $R(V_g)$ curve is too large to measure, therefore, $E_g$ cannot be quantified using Eqn. 1. We adopt a different approach. To determine $E_g$ in the dielectric liquid, we take inspiration from previous work on liquid-gated CNTs[25] and liquid-gated $WSe_2$ devices.[26] We use a conducting liquid to gate our devices and determine the gate-voltage window where the CNT is depleted of charge carriers. This gate voltage window is then equated to $E_g$. The liquid gate medium is chosen to ensure that (1) The capacitive coupling between CNT and the gate is very strong, and (2) The Schottky barriers are sufficiently thin that the onset of p-type and n-type doping is observable. To determine $E_g$ in air we utilize chiral index identification and literature values for the lowest optical resonance and the exciton binding energy, $E_b$.

Measurements of a semiconducting CNT are illustrated in Figure 6. We first determine the chiral index of the CNT by measuring the photocurrent spectrum in air (Fig. 6a & b). From the photocurrent spectrum we identify the exciton resonances and compare to the exciton resonances listed in the CNT atlas.[27] We find excellent agreement between our measured resonances and the expected resonances for chiral index (20, 18). From the chiral index we know that the lowest energy exciton resonance (in air) is 400 meV.[27] We can also estimate the exciton binding energy, $E_b$ = 130 meV, based on the results of two-photon optical characterization of chirality-enriched CNT films.[28] These two-photon experiments were performed in a dielectric environment, $\varepsilon_{env} \sim 3$, therefore 130 meV is a lower bound for $E_b$ in air. We conclude that $E_{g,air} > 0.53$ eV.

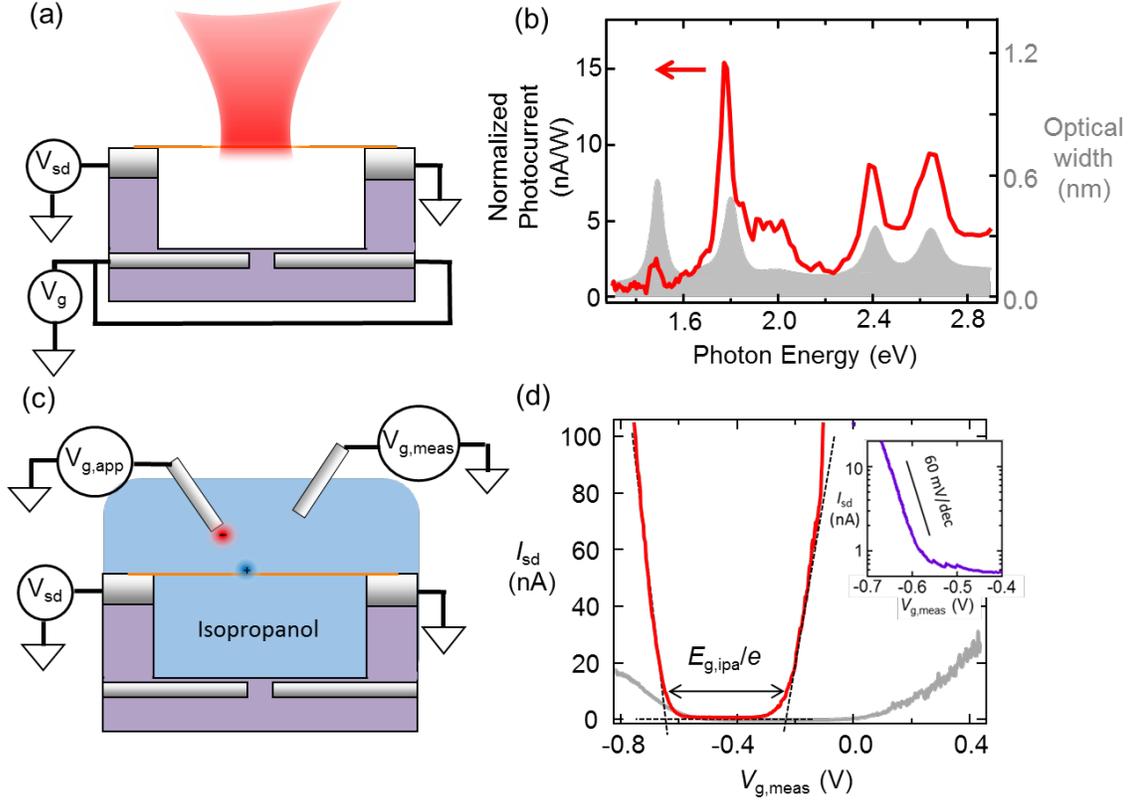

**Figure 6.** Method of measuring the change in transport gap for wide-bandgap CNTs. a) Schematic of the photocurrent spectroscopy measurement that is used to determine chiral index. b) The spectral dependence of the photocurrent (red line) plotted together with the expected absorption cross section of a (20,18) CNT. c) Schematic of a liquid-gating measurement. d) $I_{sd}$-$V_g$ in pure IPA (gray line) and salty IPA, 30mM of TBAF, (red line). Inset: Log plot of the sub-threshold region using salted IPA.

Next, the CNT is submerged in isopropanol and we measure $G(V_g)$. The gate voltage is applied to the liquid, rather than to underlying gate electrode. We are careful to minimize leakage currents between the liquid and the source/drain electrodes (see Methods). Isopropanol is used for the liquid dielectric, rather than oil, because ionic species can be dissolved in the isopropanol. Dissolved ions are critical for increasing the gate capacitance, and shrinking the length of the Schottky barriers.[26] Tetrabutylammonium fluoride (TBAF) is used as the dissolved ionic salt. The liquid potential is measured by an independent electrode via a high-impedance voltmeter (see Methods).[26] The light grey curve in Figure 6d is measured in pure isopropanol and the red curve is measured in salted isopropanol (30 mM TBAF).

To verify that the Schottky barriers are sufficiently transparent, and the liquid gate capacitance is sufficiently strong, we measure the sub-threshold slope of the $G(V_g)$ curve (see inset to Fig. 6d). In pure isopropanol we observe a sub-threshold swing ~ 120 mV/dec. In salted isopropanol the sub-threshold swing is 60 mV/dec, the lowest possible value. A sub-threshold swing of 60 mV/dec indicates that there is a one-to-one relationship between the $eV_g$ and the Fermi energy in the CNT when the Fermi energy is within the transport gap.

The transport gap is determined by extrapolating the linear regions of the $G(V_g)$ curve to the $V_g$ axis.[26] The x-intercepts define a gate-voltage window of 0.42 V, suggesting that $E_g$ = 0.42 eV. We conclude that the isopropanol environment reduces $E_g$ from > 0.53 eV to 0.42 eV. We repeated this experiment on two additional wide-gap CNTs (see Table 1). The average reduction from the lower bound value of $E_{g,air}$, to $E_{g,ipa}$ is 25%.

| Device # | (n, m)  | d (nm) | $E_{g,air}$ (meV) | $E_{g,ipa}$ (meV) |
|----------|---------|--------|-------------------|-------------------|
| 12       | (20,18) | 2.58   | > 530             | 420               |
| 13       | (27,14) | 2.83   | > 540             | 380               |
| 14       | (26,10) | 2.52   | > 600             | 440               |

**Table 1.** Modifying the transport gap of wide-gap CNTs. The transport gap was first characterized in air, $E_{g,air}$, and then in isopropanol, $E_{g,ipa}$. All devices have $L = 2$ μm.

## Discussion

Our experiments in which the environment is changed from air and isopropanol reveal a 25% reduction in $E_g$ for semiconducting CNTs and a 32% reduction in $E_g$ for narrow-gap CNTs. Comparing these percentages to theory (Fig 1a), we conclude that our observations are consistent with $\epsilon_{eff}$ increasing from about $2\epsilon_0$ to $5\epsilon_0$. These values of $\epsilon_{eff}$ can be understood as follows. In the air environment, there is a small amount of screening from the CNT itself. Previous authors have estimated $\epsilon_{eff} \sim 2\epsilon_0$ for a CNT surrounded by air or vacuum.[18]. In the isopropanol environment, additional screening comes from the dielectric liquid. The dielectric response of the liquid is frequency dependent. At low frequency, $\epsilon_{env} = 18\epsilon_0$ and at high frequencies (visible light) $\epsilon_{env} = 1.9\epsilon_0$. The relevant frequencies associated with the e-e interactions that determine band gap renormalization are currently unknown. However, from our estimate of $\epsilon_{eff}$ in the isopropanol environment ($5\epsilon_0$), we postulate that GHz or THz frequencies are most relevant to the band gap renormalization process.

Band gap renormalization calculations (Fig. 1a) and our experiments both suggest that e-e interactions have a larger effect on narrow-gap CNTs than semiconducting CNTs.

The renormalized band gap can be written as $E_g = E_{g,bare} + \Sigma$, where $\Sigma$ describes the contribution of e-e interactions (the self energy). The scaling behavior of $\Sigma$ is described by $E_{g,bare} \cdot f(E_C/E_K)$, where the dimensionless factor $f$ increases as a function of effective Coulomb energy, $E_C = e^2/\varepsilon_{eff} D$, divided by typical kinetic energy $E_K$.[29] Typical kinetic energy scales with the bare band gap and is therefore much smaller in narrow gap CNTs. Thus narrow-gap CNTs have a larger factor $f(E_C/E_K)$. This insight explains the larger enhancement factors for narrow gap CNTs (Fig. 1a), and the greater sensitivity of narrow gap CNTs to dielectric screening.

The transport gap, $E_g$, includes a Coulomb charging energy component, $e^2/C_{tot}$, where $C_{tot}$ is the total capacitance between the CNT and the nearby metal electrodes (source, drain, and gate). This charging energy is reduced when a suspended CNT device is submerged in dielectric liquid. Therefore, it is important to consider whether a change in Coulomb charging energy can account for the observed change in $E_g$. Coulomb blockade spectroscopy measurements reveal that the typical charging energy for our CNTs is no more than 15 meV (see Fig. 4 and Supporting Information). Submerging a device in refractive index oil may reduce this charging energy by a factor of 3. Thus, a narrow-gap CNT with initial $E_g = 200$ meV would be reduced to $E_g = 190$ meV in refractive index oil. Figure 5b shows that the experimentally observed changes in $E_g$ are much larger. Therefore, the observed reduction in $E_g$ cannot be explained by a simple change in charging energy.

In conclusion, we have demonstrated the experimental ability to tune the e-e interaction strength in CNTs while simultaneously monitoring transport properties. The effect of e-e interactions on the transport gap is remarkably large, verifying a long-standing theoretical prediction for semiconducting CNTs. We have extended this theoretical picture to narrow-gap CNTs and experimentally demonstrate an even larger effect narrow-gap CNTs. While it is well known that CNT transport properties are exquisitely sensitive to structural variables (i.e. diameter and chiral index), our work shows that CNTs are also extremely sensitive to dielectric screening. Knowledge of this environmental sensitivity is critical for rational design of CNT devices. The strongly-interacting electron physics governing this transport gap enhancement likely affects other low-dimensional systems that are currently under investigation.[30,31]

## Methods

*Device Fabrication*

Devices with a 2-μm source-drain separation were fabricated on 4-inch Si/SiO$_2$ wafers (500 nm oxide layer) on which gate electrodes were patterned and deposited (W/Pt 5 nm/60 nm). A layer of SiO$_2$ (800 nm thickness) was used to bury the gates. The source and drain electrodes were then patterned and deposited (W/Pt 5 nm/60 nm). Reactive ion etching was used to dig a trench between the source and drain electrodes. The Pt electrodes define the edge of the trench. CNT growth catalyst (1 nm Ti / 20 nm SiO2 / 1 nm Fe) was deposited on top of the source and drain electrodes. CNTs were grown using chemical vapor deposition in a tube furnace at 800C. The chips were shuttled in to avoid electrode degradation. The growth recipe consists of a 1 minute 1 SLM H$_2$ anneal followed by a 5 minute growth phase with 0.15 SLM ethanol, 0.3 SLM methanol, and 0.45 SLM H$_2$. The ethanol and methanol are introduced into Ar gas with a bubbler. Our process routinely yields ~ 10 CNT devices per die. Devices with source-drain separation $L = 0.3$ μm, 0.7 μm, and 1.5 μm were fabricated in a similar fashion, as described in Ref. [32].

*Liquid gate measurements*

For liquid gate measurements, the appropriate ionic concentration of TBAF was determined by measuring the sub-threshold slope of the CNT FET at various TBAF concentrations, as discussed in Supplementary Note 2.

Electrical contact to the source and drain contacts on the CNT chip is made via parylene-c-coated tungsten probe needles (Microprobes for Life Science) that are positioned using x-y-z micromanipulators. The shaft of the probe needle is coated in a 3-micron-thick insulating layer of parylene-c. The insulating coating minimizes Faradaic leakage current between the source electrode, drain electrode and liquid. Only the tip of the metal needle (~ 500 μm$^2$) contacts the liquid. Voltage is applied to the liquid using a bare tungsten needle, $V_{g,app}$. A second bare tungsten needle is used to measure the liquid potential, $V_{g,meas}$ using a voltmeter with 10 GΩ input resistance. The measurement geometry is illustrated in Fig. S16. Quantification of Faradaic currents, and the relationship between $V_{g,meas}$ and $V_{g,app}$ is discussed in Supplementary Note 2.

*Numerical simulations*

The band profile of the CNT FET in Figure 2b is calculated by first solving the 2D Laplace equation without the presence of the CNT. Then, the charge density profile along the CNT is computed by means of a self-consistent cycle. Details of the computation are given in Supplementary Note 3.

## ASSOCIATED CONTENT

Coulomb blockade data and analysis methods used to create Figure 5a. Determining the room-temperature contact resistance, $R_c$, for narrow gap CNT devices. Liquid gate methods and control experiments. Liquid gate data for devices 13 and 14. Details of self-consistent electrostatics simulations (Figure 2b). This material is available free of charge via the Internet at http://pubs.acs.org.

## ACKNOWLEDGMENT

This material is based upon work supported by the National Science Foundation under Grant No. 1151369. A portion of device fabrication was carried out in the University of California Santa Barbara (UCSB) nanofabrication facility. JOI acknowledges support from the Netherlands Organization for Scientific Research (FOM/NWO).

## AUTHOR INFORMATION

**Corresponding Author**

ethan.minot@oregonstate.edu

**Author contributions**

L.A. D.R.M. G.A.S. and E.D.M. conceived and designed the experiments. L.A., D.R.M., J.O.I., and M.O. performed the experiments. A.B. performed device simulations. M.R. performed RPA-screened transport gap calculations. All authors contributed to the analysis and interpretation of results and preparation of the manuscript.